\documentclass[twocolumn,aps,showpacs]{revtex4}
\usepackage{graphicx}
\begin{document}
\title{Planar and nonplanar nucleus-acoustic solitary waves in thermally degenerate multi-nucleus plasma systems}
\author{A. A. Mamun\footnote{Also at Wazed Miah Science Research Centre,
Jahangirnagar University, Savar, Dhaka-1342, Bangladesh.} and J.  Akter}
 \affiliation{Department of Physics,
Jahangirnagar University, Savar, Dhaka-1342, Bangladesh}
\begin{abstract}
The novel thermally degenerate plasma model (based on a  system containing relativistically and thermally degenerate inertial-less electron species,   non-relativistically and thermally degenerate inertial light nucleus species, and stationary heavy nucleus species) is considered. The basic features of planar and nonplanar solitary structures associated with the thermally degenerate pressure driven nucleus-acoustic waves propagating  in such a  thermally degenerate plasma system has been investigated. The reductive perturbation method, which is valid for small amplitude solitary waves, is used.  It is found that the effects of nonplanar cylindrical and spherical geometries, non and ultra-relativistically degenerate electron species,  thermal and degenerate pressures of electron and light nucleus species, and  number densities of light and heavy nucleus species significantly modify the basic features (viz.  speed, amplitude,  and width) of the solitary potential structures associated with thermally degenerate pressure driven nucleus-acoustic waves.  The degenerate plasma model under consideration  is so general and realistic that it is applicable not only in astrophysical compact objects like hot white dwarfs,  but also in space plasma systems like mesospheres containing positively charged heavy particles in addition to electron and ion plasma species.
\end{abstract}
\date{\today}
\maketitle
\section{Introduction}
\label{Intro}
The degenerate plasmas are defined as plasmas of so high density that the degenerate pressure (generated due to Heisenberg's uncertainty principle with infinitesimally small uncertainty in position and infinitely large uncertainty in  momenta of degenerate plasma species) cannot be  neglected in any way  \cite{Fowler26,Chandrasekhar31a,Chandrasekhar31b,Chandrasekhar36,Koester90,Horn91,Koester02,Shukla11a,Shukla11b,Brodin16}.
They usually contain degenerate electron species  \cite{Fowler26,Chandrasekhar31a,Chandrasekhar31b,Chandrasekhar36,Koester90,Horn91,Koester02,Shukla11a,Shukla11b,Brodin16}, inertial light nucleus  species (viz. ${\rm ~^{1}_{1}H}$  \cite{Fowler26,Chandrasekhar31a,Chandrasekhar31b,Chandrasekhar36}  or
${\rm ~^{4}_{2}He}$  \cite{Koester90,Horn91,Koester02} or  ${\rm ~^{12}_{~6}C}$ \cite{Koester90,Horn91,Koester02}) species,  and  heavy nucleus  species (viz. ${\rm ~^{56}_{ 26}Fe}$ \cite{Vanderburg15} or ${\rm ~^{85}_{ 37}Rb}$ \cite{Witze14} or ${\rm ~^{96}_{42}Mo}$ \cite{Witze14}). The degenerate pressure ${\cal P}_{dj}$ exerted by the degenerate plasma species $j$ (where $j=e$ for the electron species and $j=l$ for the light nucleus species) is \cite{Chandrasekhar31b,Chandrasekhar36,Koester90,Horn91,Koester02,Shukla11a,Shukla11b}
\begin{eqnarray}
{\cal P}_{dj}={\cal K}_jN_j^{\gamma_j},
\label{a}
\end{eqnarray}
where $N_j$ is the  number density of the plasma species $j$,  and  $\gamma_j$ and ${\cal K}_j$ are
\begin{eqnarray}
\gamma_j=\frac{5}{3}, ~~~
{\cal K}_j\simeq\frac{3\pi\hbar^2}{5m_j}
\label{a1}
\end{eqnarray}
for the non-relativistic limit \cite{Chandrasekhar36,Koester90,Horn91,Koester02,Shukla11a,Shukla11b}, and
\begin{eqnarray}
\gamma_j=\frac{4}{3}, ~~~
{\cal K}_j\simeq\frac{3}{4}\hbar c,
\label{a2}
\end{eqnarray}
for the ultra-relativistic  limit \cite{Chandrasekhar36,Koester90,Horn91,Koester02,Shukla11a,Shukla11b},  where  $m_j$ is the mass of the plasma species $j$,  $\hbar$ is the reduced Plank constant, and $c$ is the speed of light in vacuum.

The concept of the ion-acoustic waves \cite{Tonks29ia-theor,Revans33ia-expt} leads Mamun  \cite{Mamun18} to  introduce the electron degenerate energy $E_{e0}={\cal K}_eN_{e0}^{\gamma_e-1}$  along with the corresponding  wave speed  $C_l\,[=(E_{e0}/m_l)^{1/2}]$ and wave length-scale $L_q\,(=C_l/\omega_{pl})$ [where $\omega_{pl}= (4\pi N_{l0}Z_l^2e^2/m_l)^{1/2}$,  $Z_l$ being the charge state of the light nucleus species, $N_{e0}$ ($N_{l0}$) being the number density of the electron (light nucleus) species at equilibrium].  We note that $E_{e0}$, $C_l$ and $L_q$ are  associated with the equilibrium electron degenerate pressure,  and that $N_{e0}=Z_lN_{l0}+Z_hN_{h0}$ at equilibrium, where $N_{h0}$ ($Z_h$)  is the number density 
(charge state) of the heavy nucleus species.   Mamun  \cite{Mamun18} has then predicted the degenerate pressure driven (DPD) nucleus-acoustic (NA) waves in degenerate plasma system containing cold non or ultra-relativistically degenerate electron species
 \cite{Chandrasekhar31a,Chandrasekhar31b,Chandrasekhar36},  non-degenerate light nucleus species (viz.  ${\rm ~^{1}_{1}H}$ \cite{Chandrasekhar31a,Chandrasekhar31b,Chandrasekhar36}, or  ${\rm ~^{4}_{2}He}$ \cite{Horn91} or ${\rm ~^{12}_{~6}C}$ \cite{Koester02}, and stationary heavy nucleus species (viz. ${\rm ~^{56}_{ 26}Fe}$ \cite{Vanderburg15} or ${\rm ~^{85}_{ 37}Rb}$ \cite{Witze14} or ${\rm ~^{96}_{42}Mo}$ \cite{Witze14}).  The linear dispersion relation for
the DPD NA waves \cite{Mamun18} is  
\begin{eqnarray}
\omega=\sqrt{\frac{\gamma_e}{1+\mu}}\frac{k C_l}{\sqrt{1+\frac{\gamma_e}{1+\mu} k^2L_q^2}},
\label{NA-disp}
\end{eqnarray}
where $\mu=Z_hN_{h0}/Z_lN_{l0}$.  The linear dispersion relation (\ref{NA-disp}) can be interpreted as follows: 
\begin{itemize}
\item{The DPD NA waves are driven by the electron degenerate pressure and the DPD NA waves exist in any degenerate plasma system at absolute zero temperature, for which the DPD NA waves are completely  different from the ion-acoustic waves 
\cite{Tonks29ia-theor,Revans33ia-expt}.}
\item{The dispersion relation (\ref{NA-disp}) for the long-wavelength DPD NA waves ($kL_q\ll 1$) is given by
 \begin{equation}
 \omega = \sqrt{\frac{\gamma_e}{1+\mu}}\,kC_l.
 \label{NA-Ldisp1}
\end{equation}
}
\item{The dispersion relation (\ref{NA-Ldisp1}) for $\mu=0$ becomes
\begin{equation}
\omega = \sqrt{\gamma_e}kC_l\,,
 \label{NA-Ldisp2}
\end{equation}
which means that in DPD NA waves, the light nucleus mass density (electron degenerate pressure) provides the inertia (restoring force).}
\item{The phase speed ($\omega/k$)  of the long-wavelength DPD NA waves decreases with the rise of the value of $\mu$,  i.e.  with the number density ($N_{h0}$), and charge state ($Z_h$)  of the stationary heavy nucleus species for fixed light nucleus number density ($N_{l0}$).}  
\item{The length and time scales of the DPD NA waves \cite{Mamun18} are far different from those of  the ion-acoustic waves, since their restoring forces and the mediums of their  propagation are completely different.}
\end{itemize}

Recently,  a number of theoretical  investigations \cite{Mamun16,Mamun17,Zaman17,Zaman18,Jannat18a,Jannat18b,Sultana18a,Sultana18b,Chowdhury18,Das19}
have been made on the nonlinear propagation of the NA waves in degenerate plasma systems.   Mamun {\it et al.}  \cite{Mamun16,Mamun17} considered the strongly \cite{Mamun16} and  weakly \cite{Mamun17} coupled heavy nucleus fluid, and studied the small amplitude heavy nucleus-acoustic shock  \cite{Mamun16}  and solitary \cite{Mamun17} waves in an absolutely cold degenerate electron-light nucleus plasma system. These  shock and solitary waves \cite{Mamun16,Mamun17}  are associated with the dynamics of the  heavy nucleus species, but not with that of the light nucleus species.  Zaman \cite{Zaman17,Zaman18} have examined the effects of nonplanar geometry on the NA shock \cite{Zaman17} and solitary \cite{Zaman18}  waves propagating in absolutely cold degenerate electron-nucleus plasma systems.  On the hand,  Jannat \& Mamun \cite{Jannat18a,Jannat18b} considered an absolutely cold  degenerate plasma system containing cold degenerate electron species,  cold dissipative \cite{Jannat18a} and non-dissipative \cite{Jannat18b} light nucleus species,  and studied the effects of the stationary heavy nucleus species on the NA shock  \cite{Jannat18a}  and solitary \cite{Jannat18b} waves.  Sultana  \&  Schlickeiser \cite{Sultana18a} considered an absolutely cold multi-nucleus degenerate  plasma with relativistically degenerate electron species to study  the arbitrary amplitude NA solitons in an absolutely cold degenerate quantum plasma system.  Sultana  {\it et al.}  \cite{Sultana18b} have also investigated the modulated heavy nucleus-acoustic waves and associated rogue waves in the same quantum plasma system.   Chowdhury {\it et al.} \cite{Chowdhury18}  have studied the nucleus-acoustic envelope solitons and their modulational instability in an absolutely cold degenerate plasma system.  Das  \&  Karmakar \cite{Das19} have studied the nonlinear propagation of the heavy nucleus-acoustic waves associated with the dynamics of cold heavy nucleus species.  We note that in heavy nucleus-acoustic waves, the inertia is provided by the mass density of heavy nucleus species, but not of light nucleus species, and the length and time scales  of heavy nucleus species are far different from  that of the NA waves, which are associated with the dynamics of light nucleus species. 

The works  \cite{Mamun16,Mamun17,Zaman17,Zaman18,Jannat18a,Jannat18b,Sultana18a,Sultana18b,Chowdhury18,Das19} pinpointed above are valid only for absolutely cold degenerate plasma species, but are not valid for the thermally degenerate plasma species, particularly for hot white dwarfs \cite{Dufour08,Dufour11,Werner15,Werner19,Koester20}. To overcome the limitations of the works,  
we consider a thermally degenerate plasma (TDP).  The TDP species $j$ are governed by the  thermally degenerate pressure [defined as the sum of the degenerate pressure (${\cal P}_{dj}$) and thermal pressure (${\cal P}_{tj}$)] of the plasma species $j$.  The thermal  pressure ${\cal P}_{tj}$ of the plasma species $j$ can be expressed as
\begin{eqnarray}
{\cal P}_{tj}=N_jk_BT_j,
\label{a3}
\end{eqnarray}
where $T_j$ is the thermal temperature of TDP species $j$,  and  $k_B$ is the Boltzmann constant.   

Therefore, in our present work, we consider this novel TDP system containing thermally degenerate electron and light nucleus species, and low dense stationary heavy nucleus species, and investigate small amplitude thermally degenerate pressure driven (TDPD) NA solitary waves (SWs) by the reductive perturbation method \cite{Washimi66,Maxon74,Mamun02} for one dimensional (1D) planar 
as well as nonplanar cylindrical and spherical geometries.  The novel TDP model under our present consideration is so general that it is applicable in astrophysical compact objects like hot white dwarfs  \cite{Dufour08,Dufour11,Werner15,Werner19,Koester20} as well as in many space environments like  mesospheres containing heavy positively charged particles (as dust species) in addition to isothermal electron and inertial ion plasma species  \cite{Havnes96,Gelinas98,Mendis04}. 

The  manuscript is structured as follows. The  novel TDP model  is illustrated  in Sec. \ref{TDP-Model}. The modified  Korteweg-de-Vries (MK-dV)  equation is derived by using the reductive perturbation method,  and the basic features of the  small amplitude TDPD  NA solitary  waves are investigated  by solving  the MK-dV equation analytically for 1D planar geometry, and numerically for nonplanar cylindrical and spherical geometries in  Sec.  \ref{TDP-NA-SWs}.  The novel TDP model under consideration, results obtained from this investigation, and some important applications are  briefly discussed in Sec. \ref{Discussion}.

\section{TDP Model Equations}
\label{TDP-Model}
We consider the novel TDP system containing thermally degenerate electron and light nucleus species, and stationary heavy nucleus species.  We also consider the nonlinear propagation of TDPD NA waves in this TDP system.  The dynamics of the thermally degenerate electron species is described by the balance between the electrostatic pressure and thermally degenerate electron pressure, i.e.
\begin{eqnarray}
N_ee\frac{\partial\Phi}{\partial R}=\frac{\partial}{\partial R}({\cal P}_{de}+{\cal P}_{te}),
\label{a4}
\end{eqnarray}
where $\Phi$ is the TDPD NA wave electrostatic potential and $R$ is the space variable.  Now, substituting $P_{de}$ and $P_{te}$ 
[which are obtained from (\ref{a}) and  (\ref{a3}) for the electron species,  respectively], we obtain
\begin{eqnarray}
n_e&=&\left(1+\frac{\gamma_e-1}{\gamma_e}\phi\right)^{\frac{1}{\gamma_e-1}}, \nonumber\\
      &\simeq&\left(\frac{1}{\gamma_e}\right)\phi + \left(\frac{\gamma_e^{\prime}}{2!\gamma_e^2}\right)\phi^2+
      \left(\frac{\gamma_e^{\prime}\gamma_e^{\prime\prime}}{3!\gamma_e^3}\right)\phi^3\cdot\cdot\cdot,
\label{be1}
\end{eqnarray}
where $\gamma_e^{\prime}= 2- \gamma_e$,  $\gamma_e^{\prime\prime}=3 - 2\gamma_e$,  $n_e=N_e/N_{e0}$ and $\phi=e\Phi/{\cal E}_{e0}$ with  ${\cal E}_{e0}= E_0+k_BTe$.
It is obvious that for the cold degenerate electron species,  $T_e=0$ and ${\cal E}_{e0}=E_0={\cal K}_eN_{e0}^{\gamma_e-1}$, which mean that $\phi=e\Phi/{\cal E}_{ed}=e\Phi/E_0$.  On the other hand, for the non-degenerate thermal electron species,
$E_0=0$ and ${\cal E}_{e0}=k_BT_e$, which indicate that $\phi=e\Phi/k_BT_e$. We note that  $\gamma_e=1$
reduces the expansion form of (\ref{be1}) to $n_e=\exp(\phi)$. Thus, (\ref{be1}) is valid for arbitrary value of $\gamma_e$, and is, thus, valid  for $\gamma_e=1$ (Boltzmann distributed isothermal electron species),  $\gamma_e=5/3$  (non-relativistically thermally degenerate electron species), and  $\gamma_e=4/3$ (ultra-relativistically  thermally degenerate electron species).

The nonlinear dynamics of the TDPD NA waves in such a new TDP system in planar or nonplanar geometry
is governed by  
\begin{eqnarray}
&&\frac{\partial n_l}{\partial t}
+\frac{1}{r^\nu} \frac{\partial}{\partial r} (r^\nu n_lu_l) = 0,
\label{be2}\\
&&\frac{\partial u_l}{\partial t} +u_l\frac{\partial u_l}{\partial r}=-\frac{\partial
\phi}{\partial r} -\frac{5}{2}\beta\frac{\partial n_l^{\frac{2}{3}}}{\partial r}-\frac{\sigma_l}{n_l}\frac{\partial n_l}{\partial r},
\label{be3}\\
&&\frac{1}{r^\nu}\frac{\partial}{\partial r}\left(r^\nu\frac{\partial
\phi}{\partial r}\right) =(1+\mu)n_e-n_l-\mu,
\label{be4}
\end{eqnarray}
where $\nu=0$ for 1D planar geometry and $\nu=1\,(2)$ for
a nonplanar cylindrical (spherical) geometry;  $n_e$ is given by (\ref{be1});  $n_l$ is the light nucleus number density normalized by $N_{l0}$; 
$u_l$ is the light nucleus fluid speed normalized by $C_q\,[=({\cal E}_{e0}/m_l)^{1/2}]$;  $r=R/\lambda_q$ with
$\lambda_q=C_q/\omega_{pl}$;  $t$ is the time variable
normalized by $\omega_{pl}^{-1}$; $\sigma_l=k_BT_l/Z_{l}{\cal E}_{e0}$, and
$\beta=(m_e/Z_l m_l) (n_{l0}/n_{e0})^{2/3}$.

The normalized basic equations (\ref{be1})$-$(\ref{be4})  in 1D planar or cylindrical or spherical geometries (depending on value of $\nu$)  can be interpreted as follows: 
\begin{itemize}
\item{The equation (\ref{be1}) represents the expression for thermally degenerate inertia-less electron number density, which  is valid for the arbitrary value of $\gamma_e$.}
\item{The equation (\ref{be2}) represents the continuity equation for the thermally degenerate light nucleus species, where the effects of the source and sink terms are neglected in our present investigation.}
\item{The equation (\ref{be3}) represents the momentum balance equation for thermally degenerate light nucleus species, and 
the 2nd  (3rd) term on the right hand side of  (\ref{be2})  is due to the effect of non-relativistically degenerate (isothermal) light nucleus species.} 
\item{The equation (\ref{be4}) is Poisson's equation for the DPD NA wave potential, which closes (\ref{be1})$-$(\ref{be3}), and the parameter  $\mu$  in  (\ref{be4}) is due to  the assumption of stationary  heavy nucleus species, which is valid since the heavy nucleus plasma frequency is much less than the DPD NA wave frequency because of the heavy mass and low number density of this heavy nucleus species.}
\end{itemize}
It is important to mention here that the gravitational force acting in thermally degenerate electron and light nucleus species, which is inherently very small compared to the other forces under consideration, is neglected for this investigation. We also note that  that $\nu$ arises due to the effects of nonplanar cylindrical ($\nu=1$) and  spherical ($\nu=2$) geometries.
\section{TDPD NA Solitary Waves}
\label{TDP-NA-SWs}
To study the planar and nonplanar  DPD-NA solitary waves by the reductive perturbation method \cite{Washimi66},
we stretch the independent $x$ and $t$ as \cite{Maxon74,Mamun02}
\begin{eqnarray}
&&\xi=-\epsilon^{\frac{1}{2}}(r+{\cal M}t),
\label{str1}\\
&&\tau=\epsilon^{\frac{3}{2}}t,
\label{str2}
\end{eqnarray}
and expand the dependent variables ($n_l$, $u_l$ and $\phi$) as \cite{Washimi66}
\begin{eqnarray}
&&n_l=1+\epsilon n_l^{(1)}+\epsilon^2 n_l^{(2)}+\cdot \cdot \cdot,\\
&&u_l=0+\epsilon u_l^{(1)}+\epsilon^2 u_l^{(2)}+\cdot \cdot \cdot,\\
&&\phi=0+\epsilon \phi^{(1)}+\epsilon^2 \phi^{(2)}+\cdot \cdot \cdot,
\label{expansion}
\end{eqnarray}
where ${\cal M}$ is the phase speed (normalized by $C_q$) of the DPD-NA waves (i.e.  ${\cal M}=\omega/kC_q$), and $\epsilon$ is the stretching or expansion parameter with a value satisfying $0<\epsilon<1$.

Now, using (\ref{str1})$-$(\ref{expansion}) in (\ref{be1})$-$(\ref{be4}), and taking the coefficients of $\epsilon^{3/2}$ from  (\ref{be2}) and (\ref{be3}), and the coefficients of $\epsilon$ from (\ref{be4}), we obtain
\begin{eqnarray}
&&n_l^{(1)}=\frac{1}{{\cal M}^2-\beta_\sigma}\phi^{(1)},
\label{1st-order1}\\
&&u_l^{(1)}=\frac{{\cal M}}{{\cal M}^2-\beta_\sigma}\phi^{(1)},
\label{1st-order2}\\
&&{\cal M}=\bigg[\frac{\gamma_e+(1+\mu)\beta_\sigma}{1+\mu}\bigg]^{\frac{1}{2}},
\label{1st-order3}
\end{eqnarray}
where $\beta_\sigma=\sigma_l+ 5\beta/3$. We note that (\ref{1st-order3}) represents the linear dispersion relation for the TDPD NA waves. This linear dispersion relation can be written in dimensional form as
\begin{eqnarray}
\omega=kC_q\sqrt{\frac{\gamma_e+(1+\mu)\beta_\sigma}{1+\mu}}.
\label{L-disp}
\end{eqnarray}
It is obvious that for non-degenerate cold light nucleus species, $\beta=0$, 
$\sigma_l=0$,  $C_q=C_l$, and the dispersion relation ({\ref{L-disp}}) is identical to (\ref{NA-Ldisp1}). This means that the linearized equations (\ref{1st-order1})$-$(\ref{1st-order3}), which are derived by using the 1st order approximation, are valid for the long wavelength TDPD NA waves.

Similarly, using (\ref{str1})$-$(\ref{expansion}) in (\ref{be1})$-$(\ref{be4}), and taking the coefficients of $\epsilon^{5/2}$ from (\ref{be2}) and (\ref{be3}), and the coefficient of $\epsilon^2$ from (\ref{be4}),  we obtain
\begin{eqnarray}
\hspace*{-9mm}&&\frac{\partial n_l^{(1)}}{\partial\tau}-\frac{\partial}{\partial\xi}\left[{\cal M}n_l^{(2)}+ u_l^{(2)}+n_l^{(1)}u_l^{(1)}\right]-\frac{\nu u_l^{(1)}}{\tau{\cal M}}=0, 
\label{2nd-order1}\\
\hspace*{-9mm}&&\frac{\partial u_l^{(1)}}{\partial\tau}-{\cal M}\frac{\partial u_l^{(2)}}{\partial\xi}-u_l^{(1)}\frac{\partial u_l^{(1)}}{\partial\xi}=\frac{\partial \phi^{(2)}}{\partial\xi}+\frac{5}{3}\beta\frac{\partial n_l^{(2)}}{\partial\xi}\nonumber\\
\hspace*{-9mm}&&~~~~~~~~~~~~~~~~~~~~~~~~~+\frac{1}{2}\left(\sigma_l- \frac{5}{9}\beta\right)\frac{\partial}{\partial\xi}\left[n_l^{(1)}\right]^2,
\label{2nd-order2}\\
\hspace*{-9mm}&&\frac{\partial^2\phi^{(1)}}{\partial\xi^2}=\left(\frac{1+\mu}{\gamma_e}\right)\phi^{(2)}+\frac{(1+\mu)\gamma_e^\prime}{2\gamma_e^2}\left[\phi^{(1)}\right]^2-n_l^{(2)},
\label{2nd-order3}
\end{eqnarray}
where $\beta_\sigma^\prime=(\sigma_l- 5\beta/9)/2$. Using (\ref{1st-order1})$-$(\ref{2nd-order3}), we eliminate $\phi^{(2)}$, $u_l^{(2)}$ and $n_l^{(2)}$  to obtain
\begin{eqnarray}
\hspace*{-9mm}&&\frac{\partial \phi^{(1)}}{\partial \tau} +\frac{\nu}{2\tau}\phi^{(1)} + {\cal A}\phi^{(1)} \frac{\partial
\phi^{(1)}}{\partial \xi} +{\cal B} \frac{\partial^{3} \phi^{(1)}}{\partial \xi^{3}}=0,
\label{MKdV}
\end{eqnarray}
where ${\cal A}$ and ${\cal B}$ are nonlinear and dispersion coefficients,
respectively, and are  given by
\begin{eqnarray}
\hspace*{-7mm}&&{\cal A}={\cal B}\left(\frac{\mu_1}{\gamma_e}\right)^2\left[3+2\frac{\mu_1}{\gamma_e}\left(\sigma_l+\frac{20}{9}\beta\right)+\frac{\gamma_e^\prime}{\mu_1}\right],
\label{coffA}\\
\hspace*{-7mm}&&{\cal B} =\frac{\gamma_e^2}{2\mu_1^{\frac{3}{2}}\sqrt{\gamma_e+\mu_1\beta_\sigma}},
\label{coffB}
\end{eqnarray}
where $\mu_1=1+\mu$. We note that (\ref{MKdV}) is the MK-dV equation with the nonlinear coefficient ${\cal A}$ defined by (\ref{coffA}), and the dispersion coefficient $B$ defined by (\ref{coffB}), and that ${\cal A}>0$ and  ${\cal B}>0$ for all  possible values of  $\gamma_e$, $\mu$, $\sigma_l$, 
and $\beta$.  To avoid any confusion, we also note that $1\ge\gamma_e<2$ since $\gamma_e=1$  for isothermal electron species, $\gamma_e=5/3$  for non-relativistically degenerate electron species, and $\gamma_e=4/3$ for ultra-relativistically degenerate electron species, which are of our present interest.  It is important to mention here that that the 2nd term ($\nu\phi^{(1)}/2\tau$) in (\ref{MKdV}) is due to the effect  of the nonplanar geometry, and that for $\nu=0$ or $\tau\rightarrow \infty$,  the effect  of the nonplanar geometry disappears, and turns to be the 1D planar geometry.
\subsubsection{1D Planar geometry ($\nu=0$)}
We first consider planar geometry by substituting $\nu=0$ or $\tau\rightarrow \infty$ into (\ref{MKdV}). The assumption  $\nu=0$ or $\tau\rightarrow \infty$ reduces (\ref{MKdV}) to a standard K-dV equation. To obtain the stationary solitary wave (SW) solution of the standard  K-dV equation or (\ref{MKdV}) with $\nu=0$, we consider a moving frame
$\zeta=\xi-{\cal U}_0\tau$, where ${\cal U}_0$ is the SW speed in the moving frame,
and  impose the appropriate boundary conditions, viz.  $\phi^{(1)}\rightarrow 0$ and
$d\phi^{(1)}/d\zeta\rightarrow 0$ at $\zeta\rightarrow \pm \infty$. Thus, the stationary SW solution of  (\ref{MKdV}) with $\nu=0$  
is given by
\begin{eqnarray}
&&\phi^{(1)}=\psi\, {\rm sech}^2\left(\frac{\zeta}{\delta}\right),
\label{solution}
\end{eqnarray}
where $\psi$ and $\delta$, respectively, represent the amplitude and the width of the TDPD NA SWs, since (\ref{solution}) represents the TDPD NA SWs with the positive electrostatic wave potential, corresponding to the density compression (in the form of the solitary profile) of the light nucleus species. This is because of ${\cal A}>$ and ${\cal B}>0$.  The amplitude ($\psi$) and the width ($\delta$) of the TDPD NA SWs are, respectively, given by
\begin{eqnarray}
&&\psi =\frac{3 {\cal U}_0}{{\cal A}},
\label{amp}\\ 
&&\delta=\sqrt{\frac{4 {\cal B}}{{\cal U}_0}}.
\label{wid}
\end{eqnarray}
It is obvious from (\ref{amp}) and (\ref{wid}) that the amplitude (width) of the TDPD NA SWs directly (inversely) proportional to ${\cal U}_0$ (i.e. $\psi\propto {\cal U}_0$ and $\delta\propto 1/{\cal U}_0$).  

\begin{figure*}
\centering
\begin{tabular}{@{}cc@{}}
\includegraphics[width=0.48\textwidth]{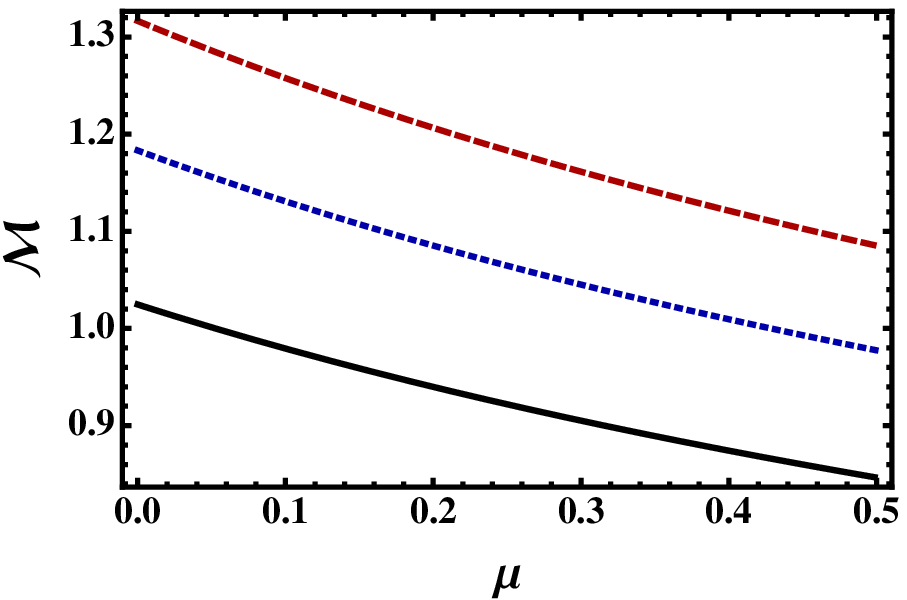} &
\includegraphics[width=0.48\textwidth]{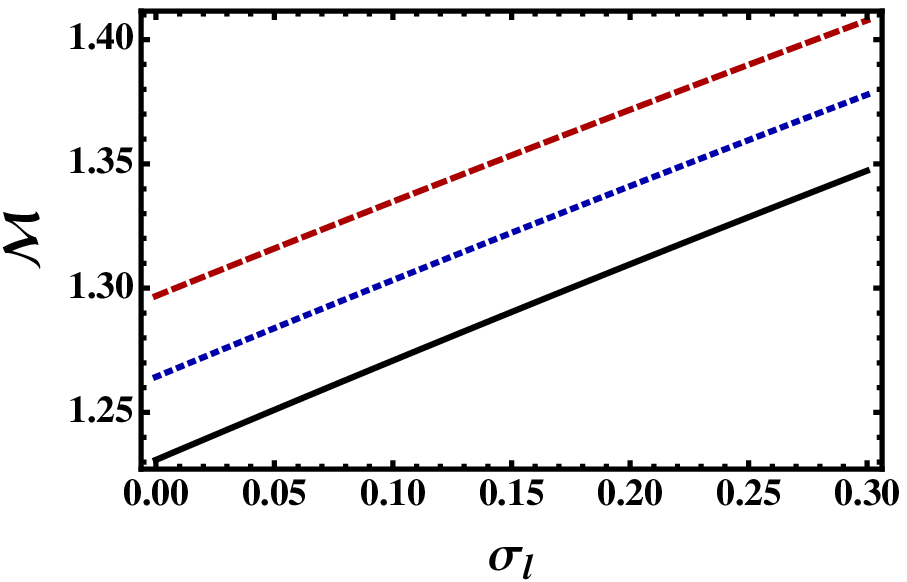}
\end{tabular}
\caption{The variation of the normalized TDPD NA wave phase speed ${\cal M}$ with $\mu$ for 
$\sigma_l=0.05$, \{$\beta=0.0$, $\gamma_e=1$\} (solid curve), \{$\beta=0.01$,$\gamma_e=4/3$\} (dotted curve) and  \{$\beta=0.01$,$\gamma_e=5/3$\} (dashed curve) in the left panel, whereas that with $\sigma_l$ for  $\gamma_e=5/3$, $\mu=0.1$,  $\beta=0$ (solid curve), $\beta=0.05$ (dotted curve) and  
$\beta=0.1$ (dashed curve) in the right panel.}
\label{f1-f2-M}
\end{figure*}

\begin{figure*}
\centering
\begin{tabular}{@{}cc@{}}
\includegraphics[width=0.48\textwidth]{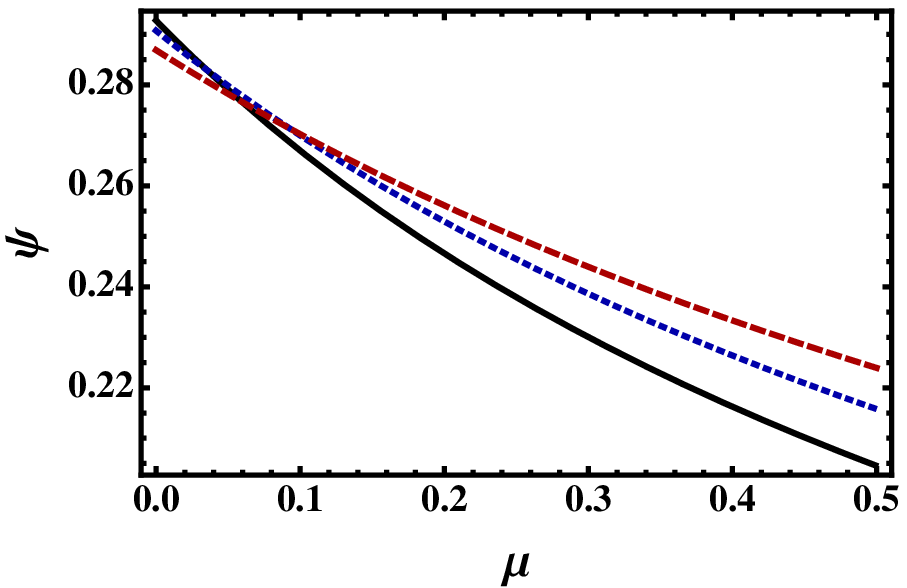} &
\includegraphics[width=0.48\textwidth]{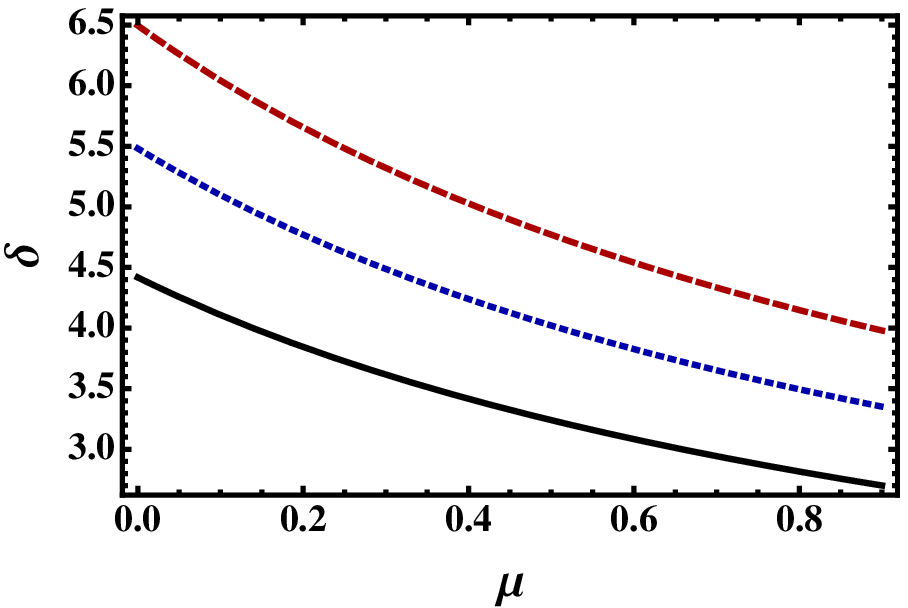}
\end{tabular}
\caption{The variation of the normalized TDPD NA SW amplitude $\psi$ (left panel) and width $\delta$ 
(right panel)  with  $\mu$  for $\sigma_l=0.05$, ${\cal U}_0=0.1$, \{$\beta=0.0$, $\gamma_e=1$\} (solid curve), \{$\beta=0.01$,$\gamma_e=4/3$\} (dashed curve).}
\label{f3-f4-amp-width}
\end{figure*}

\begin{figure*}
\centering
\begin{tabular}{@{}cc@{}}
\includegraphics[width=0.48\textwidth]{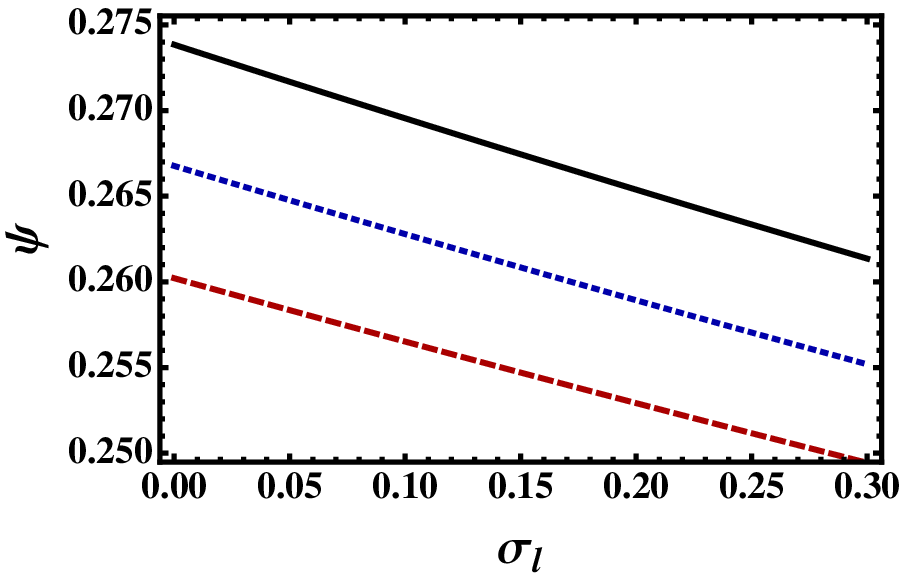} &
\includegraphics[width=0.48\textwidth]{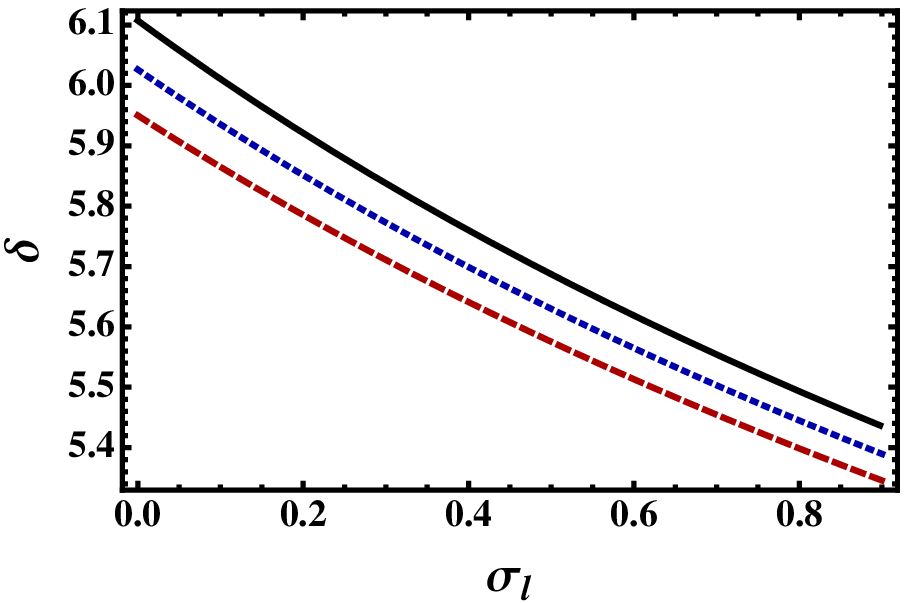}
\end{tabular}
\caption{The variation of the normalized TDPD NA SW amplitude $\psi$ (left panel)  and width $\delta$ 
(right panel) with $\sigma_l$ for $\mu=0.1$, $\gamma_e=5/3$,  ${\cal U}_0=0.1$,  $\beta=0$ (solid curve), 
$\beta=0.05$ (dotted curve), and
$\beta=0.1$ (dashed curve).}
\label{f5-f6-amp-width}
\end{figure*}

We now graphically show how the basic features, viz.  speed ${\cal M}$,  amplitude $\psi$, and width $\delta$ of the  TDPD NA SWs vary with typical plasma parameters, viz.  $\mu$,  $\gamma_e$, $\sigma_l$  and $\beta$ corresponding  not only to hot white dwarfs  \cite{Dufour08,Dufour11,Werner15,Werner19,Koester20}, but also to many space plasma environments, where non-degenerate electron-ion plasma with heavy positively charged particles (as impurity or dust) occur \cite{Havnes96,Gelinas98,Mendis04}.  
The results are displayed in figures \ref{f1-f2-M}$-$\ref{f5-f6-amp-width}, and are interpreted in discussion section.  It is important to mention here that for completely non-degenerate plasma system (containing isothermal electron species,  non-degenerate warm ion species, and positively charged stationary heavy particles  as impurity or dust species), $\gamma_e=1$ and $\beta=0$ (shown in the solid curve in the left panel of figure \ref{f1-f2-M},  whereas in both left and right panel of figure \ref{f3-f4-amp-width}, and that these solid curves correspond to the space plasma environments like mesosphere \cite{Havnes96,Gelinas98,Mendis04}. 
\subsubsection{Non-planar geometry ($\nu\ne0$)}
\begin{figure*}
\centering
\begin{tabular}{@{}cc@{}}
\includegraphics[width=0.48\textwidth]{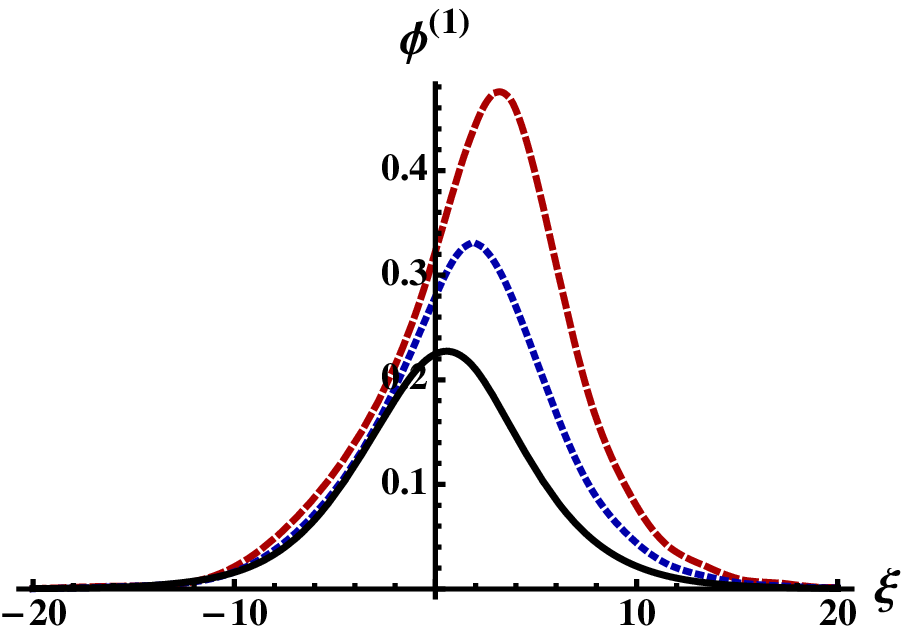} &
\includegraphics[width=0.48\textwidth]{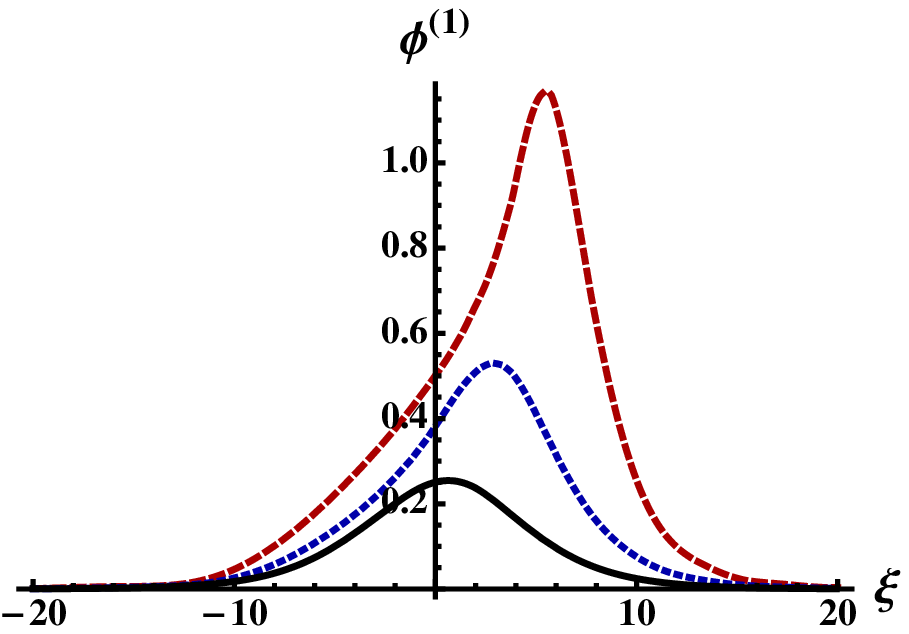}
\end{tabular}
\caption{The effects of cylindrical (left panel) and spherical (right panel) geometries on the  SWs associated with the TDPD NA wave potential for  ${\cal U}_0=0.1$,  $\mu=0.3$,  $\sigma_l=0.1$,  $\gamma_e=5/3$, $\beta=0.05$,  $\tau=-16$ (solid curves),  $\tau=-8$ (dotted curves) and  $\tau=-4$ (dashed curve).}
\label{f7-f8-cyl-sph}
\end{figure*}

We are now interested in direct numerical solutions of (\ref{MKdV}) to observe the time evolution of the TDPD NA cylindrical ($\nu=1$) and spherical ($\nu=2$) SWs in the TDP systems under consideration.  The initial pulse that we  have used in our numerical analysis is  (\ref{solution}) which valid for 
$\tau\rightarrow \pm \infty$.  To mean `time from past to present', we use $\tau<0$.  Thus, the  initial pulse for the numerical solutions of  (\ref{MKdV} can be expressed as 
\begin{eqnarray}
&&\phi^{(1)}(\tau\rightarrow\pm\infty)=\psi\, {\rm sech}^2\left(\frac{\zeta}{\delta}\right).
\label{solution-t}
\end{eqnarray}
We note that for a large value of $\tau$, the TDPD NA solitary profiles represented by (\ref{solution}) are identical to those represented by (\ref{solution-t}), since the effects of nonplanar geometry represented by the term  $\nu\phi^{(1)}/2\tau$ disappear for the large value of $\tau$. We have obtained the direct numerical solutions of (\ref{MKdV}) using this initial pulse, and using typical plasma parameters corresponding to hot white dwarfs  \cite{Dufour08,Dufour11,Werner15,Werner19,Koester20}. The direct numerical solutions of (\ref{MKdV}) is displayed in figure \ref{f7-f8-cyl-sph}.

\begin{figure*}
\centering
\begin{tabular}{@{}cc@{}}
\includegraphics[width=0.48\textwidth]{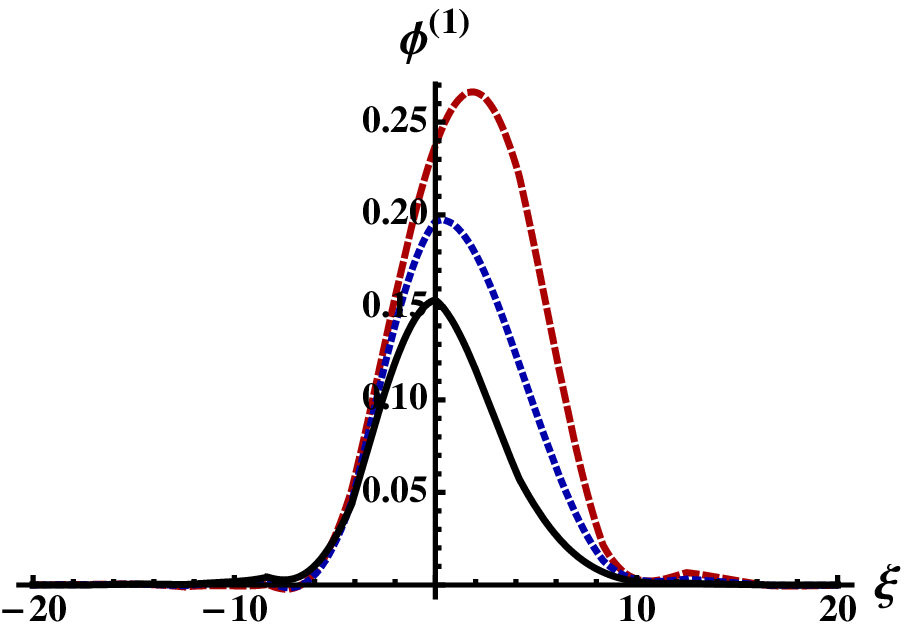} &
\includegraphics[width=0.48\textwidth]{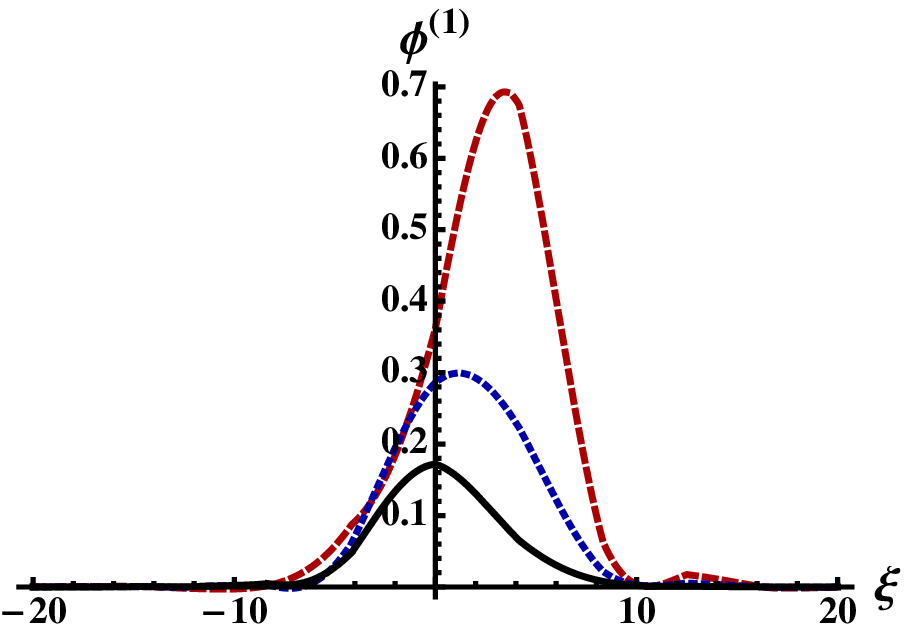}
\end{tabular}
\caption{The effects of cylindrical (left panel) and spherical (right panel) geometries on the  SWs associated with the TDPD NA wave  potential for  ${\cal U}_0=0.1$,  $\mu=0.3$,  $\sigma_l=0.1$,  $\gamma_e=1$, $\beta=0$,  $\tau=-16$ (solid curves),  $\tau=-8$ (dotted curves and  $\tau=-4$ (dashed curve).}
\label{f9-f10-cyl-sph}
\end{figure*}

We have also found the direct numerical solutions of (\ref{MKdV}) using the initial pulse represented by  (\ref{solution-t}), and using typical plasma parameters corresponding to space plasma environments like mesospheres, where $\gamma_e=1$ and $\beta=0$  \cite{Havnes96,Gelinas98,Mendis04}.
The direct numerical solutions of (\ref{MKdV}) is displayed in figure \ref{f9-f10-cyl-sph}.

The direct numerical solutions of (\ref{MKdV}), which have been displayed in figures \ref{f7-f8-cyl-sph} and \ref{f9-f10-cyl-sph}, clearly indicate that for a large value of $\tau$ (e.g. $\tau=-16$) the spherical and cylindrical TDPD NA SWs are similar to the 1D planar ones. This is because that the time and geometry dependent extra term $(\nu/2\tau)\phi^{(1)}$  in (\ref{MKdV}) has become insignificant.  However, as the value of $\tau$ decreases, the term
$(\nu/2\tau)\phi^{(1)}$ becomes significant, and both the spherical and cylindrical SWs differ from the 1D planar ones.
\section{Discussion}
 \label{Discussion}
The novel TDP model (based on the plasma system containing thermally degenerate electron species, thermally degenerate light nucleus species, and low dense stationary heavy nucleus species) has been considered to study the TDPD NA SWs  in 1D planar  as well as in cylindrical and spherical nonplanar geometries. The basic features of  these TDPD NA SWs in such a plasma system have been investigated by the reductive perturbation method \cite{Washimi66,Maxon74,Mamun02}, which is valid for small but finite  amplitude TDPD NA SWs. The results, which have been observed from this theoretical and numerical  investigations, can be summarized as follows:

\begin{enumerate}
\item{The presence of  stationary heavy nucleus species in electron-nucleus plasma supports the existence of subsonic  TDPD NA SWs with $\phi>0$ in both planar and nonplanar geometries.  The increase in number  density of stationary heavy nucleus species increases (decreases), but that in thermal and degenerate pressures  of the TDP species  decreases  (increases) the possibility for the formation of the  subsonic (supersonic) TDPD NA SWs. This is due to the fact that the phase speed of the TDPD NA waves decreases (increases) with the rise of the number density of the stationary heavy nucleus species (thermal and degenerate pressures of the TDP species).}

\item{The amplitude (width) of the TDPD NA SWs  increases (decreases) with the rise of the number density of the heavy  nucleus species in both planar and nonplanar geometries. This is because of the decrease in the TDPD NA wave phase speed with the rise of the stationary heavy nucleus  number density.}

\item{The amplitude (width) of the TDPD NA SWs  decreases (increases) with the rise of the light nucleus thermal pressure in both planar and nonplanar geometries. This is due to the fact that thermal  pressure of the light nucleus fluid  enhances the random motion of the light nucleus species, which causes to decrease (increase) the amplitude (width) of the TDPD NA solitary structures in any geometry under consideration.}

\item{The amplitude of the TDPD NA SWs for ultra-relativistically degenerate electron species  is smaller than that for non-relativistically degenerate electron species, but is larger than that for isothermal electron species. On other hand, their width for non-relativistically degenerate electron species  is  larger than that for ultra-relativistically degenerate electron species, but is smaller than that for isothermal electron species  in both planar and nonplanar geometries. This is due to that fact that the amplitude (width) of the SWs increases (decreases) as the value of $\gamma_e$ increases.}

\item{The amplitude of cylindrical and spherical TDPD NA SWs increases as the value of $\tau$ decreases for  both degenerate and non-degenerate plasmas.  The amplitude of cylindrical TDPD NA SWs is larger than that of 1D planar ones, but  smaller than that of the spherical ones. This is due to the  time and geometry dependent extra term $(\nu/2\tau)\phi^{(1)}$ in the MK-dV equation.}
\end{enumerate}

There are many hot white dwarfs \cite{Dufour08, Dufour11,Werner15,Werner19,Koester20}, where the electron thermal pressure can be comparable to or greater than its degenerate pressure, and  where in addition to non or ultra-relativistically degenerate electron species \cite{Chandrasekhar31a,Chandrasekhar31b,Chandrasekhar36,Horn91} and non-relativistically degenerate light nucleus species  (viz.  ${\rm ~^{1}_{1}H}$ \cite{Chandrasekhar31a,Chandrasekhar36}, or  ${\rm ~^{4}_{2}He}$ \cite{Horn91} or ${\rm ~^{12}_{~6}C}$ \cite{Koester02}, the stationary heavy nucleus species (viz. ${\rm ~^{56}_{ 26}Fe}$ \cite{Vanderburg15} or ${\rm ~^{85}_{ 37}Rb}$ \cite{Witze14} or ${\rm ~^{96}_{42}Mo}$ \cite{Witze14}) exists. 

On the other hand, non-degenerate electron species [defined by (\ref{be1}) as an special case of $\gamma_e=1$], ions [identical to light nucleus species], and positively charged  particle (impurity/dust species [identical to stationary heavy nucleus species] are observed in many space  plasma environments \cite{Havnes96,Gelinas98,Mendis04}.

The novel TDP model under our present consideration is so general that it is applicable not only in astrophysical compact objects like hot white dwarfs  \cite{Dufour08,Dufour11,Werner15,Werner19,Koester20}, but also in many space environments like  mesospheres \cite{Havnes96,Gelinas98,Mendis04} containing positively charged heavy stationary particles (as impurities or dust) in addition to isothermal  electron and warm ion plasma species. 

The mesospheric plasma parameters corresponds to 
$\gamma_e=1$, $\beta=0$, $\mu=0.1-0.3$, and $\sigma_l=0.01-0.1$. The basic features of the  subsonic and supersonic SWs are displayed in solid curves in left panel of figure \ref{f1-f2-M} and in both left and right panels of figure \ref{f3-f4-amp-width} for 1D planar geometry,  and in all cures in figure  \ref{f9-f10-cyl-sph} for cylindrical (left panel) and spherical (right panel)  geometries.

The limitation of the reductive perturbation method  used here is that it is not valid for arbitrary or  large amplitude TDPD NA SWs. To overcome this limitation, we have to develop a correct numerical code, and  to solve 
(\ref{be2})$-$(\ref{be4})  with (\ref{be1}) numerically by the numerical code to be developed. This type of numerical analysis  will be able to show the time evolution of large amplitude TDPD NA SWs.  This is  a challenging research problem of great importance, but beyond the scope of our present work.

However, we hope that our present work will be helpful to understand the physics of localized electrostatic disturbances in astrophysical compact objects like hot white dwarfs, e.g.  DQ white dwarfs \cite{Dufour08,Dufour11,Koester20},  white dwarf H1504+65  \cite{Werner15},  and white dwarf PG 0948+534 \cite{Werner19}. We also expect that our present work will also be useful for understanding the physics of localized electrostatic disturbances in many space environments,  e.g. mesosphere \cite{Havnes96,Gelinas98,Mendis04}.

\end{document}